\documentclass[letter]{aa}
\usepackage{graphicx}
\usepackage{txfonts}
\usepackage{caption}
\usepackage{subcaption}
\begin{document} 

   \title{A redshifted excess in the broad emission lines after the flare of the $\gamma$-ray narrow-line Seyfert 1 PKS 2004-447}

   \author{W. Hon\inst{1, 2}
          \and
          M. Berton\inst{1}
          \and
          E. Sani\inst{1}
          \and
          R. Webster\inst{2}
          \and
          C. Wolf\inst{3,4}
          \and
          A. F. Rojas\inst{5, 6}
          \and
          P. Marziani\inst{7}
          \and
          J. Kotilainen\inst{8, 9}
          \and
          E. Congiu\inst{10,1}
          }

   \institute{European Southern Observatory, Alonso de C\'ordova 3107, Casilla 19, Santiago 19001, Chile\\
              \email{whon@student.unimelb.edu.au}
         \and
             School of Physics, University of Melbourne, Parkville, Victoria 3010, Australia
        \and
            Research School of Astronomy and Astrophysics, Australian National University, Canberra ACT 2611, Australia
        \and
            Centre for Gravitational Astrophysics, Australian National University, Canberra ACT 2611, Australia
        \and
            Centro de Astronom\'{i}a (CITEVA), Universidad de Antofagasta, Avenida Angamos 601, Antofagasta, Chile
        \and
            N\'ucleo de Astronomía de la Facultad de Ingenier\'ia, Universidad Diego Portales, Av. Ej\'ercito Libertador 441, Santiago 22, Chile
        \and
            INAF - Osservatorio Astronomico di Padova, Vicolo dell'Osservatorio 5, 35122 Padova, Italy
        \and
            Finnish Centre for Astronomy with ESO (FINCA), University of Turku, Quantum, Vesilinnantie 5, 20014 University of Turku, Finland
        \and
            Department of Physics and Astronomy, University of Turku, Quantum, Vesilinnantie 5, 20014 University of Turku, Finland
        \and
            Departamento de Astronom\'{i}a, Universidad de Chile, Camino del Observatorio 1515, Las Condes, Santiago, Chile;
             }

   \date{Received September 15, 1996; accepted March 16, 1997}
   \authorrunning{WJ. Hon et al.}
   \titlerunning{Red-excess in PKS 2004-447}
    \abstract
    {PKS 2004-447 is a narrow-line Seyfert 1 (NLS1) harbouring a relativistic jet with $\gamma$-ray emission. On 2019-10-25, the \textit{Fermi}-Large Area Telescope captured a $\gamma$-ray flare from this source, offering a chance to study the broad-line region (BLR) and jet during such violent events. This can provide insights to the BLR structure and jet interactions, which are important for active galactic nuclei and host galaxy coevolution. We report X-Shooter observations of enhancements in the broad line components of Balmer, Paschen and \ion{He}{i} lines seen only during the post-flare and vanishing 1.5 years after. These features are biased redward up to $\sim$250 km s$^{-1}$ and are narrower than the pre-existing broad line profiles. This indicates a connection between the relativistic jet and the BLR of a young AGN, and how $\gamma$-ray production can lead to localised addition of broad emission lines.}
    \maketitle
    
\section{Introduction}
Since the first observations \citep{seyfert43} of broad features in AGN spectra, the nature of the emitting region in terms of its motions and structure has been the subject of much discussion and speculation \citep[e.g.,][]{mathews85, leighly07, bon09, shapovalova09, yong17}. The general view of the BLR is a region of photoionised gas that is subjected to high velocities (FWHM$>$2000 km s$^{-1}$) through random dispersion and/or virialised orbital rotational motions, as well as a potential for bulk infall or outflow \citep[e.g.,][]{wanders95, done96, denney09, williams18}. 

BLR outflows are of particular interest to the topic of AGN and host galaxy co-evolution \citep{zubovas14, king15, harrison18}. In addition to cases of blue-shifted emission lines \citep[e.g.,][]{ge19}, broad-absorption lines are observed in $\sim$20\% of quasars \citep{hamann93, trump06}, and can reach velocities up to $v\sim0.3c$ \citep{rogerson16, choi20}. This is sufficient energy \citep{scannapieco04, hopkins10, harrison18} to effectively cause feedback to the host galaxy as it eventually reaches and interacts with the interstellar medium. 

Narrow-line Seyfert 1 (NLS1) are mainly characterised by broad emission lines that are observed to be narrower than typical AGN (FWHM$<2000$ km s$^{-1}$). They also show strong \ion{Fe}{ii} multiplet emissions and low flux ratio [\ion{O}{iii}]/H$\beta < 3$ \citep[see][for an in-depth review]{komossa07}. NLS1s harbour low mass supermassive black holes of 10$^6$-10$^8$ M$_\odot$ and are considered early stage AGN\citep{boller95, peterson00, cracco16, rakshit17, chen18}. $\sim$7\% of them harbour relativistic jets \citep{mathur00, sulentic00, grupe00, komossa06}. A few dozen of them have been identified as $\gamma$-ray sources (hereafter $\gamma$-NLS1s) after the launch of \textit{Fermi}-Large Area Telescope \citep{abdo09, foschini20}. $\gamma$-NLS1s have the added properties of superluminal motion, violent variability, and optical polarisation from the relativistic jets \citep{liu10, itoh13, maune12, paliya13}. Specifically, they have very similar properties to the beamed jetted AGN sub-class of flat-spectrum radio quasars \citep[FSRQ,][]{foschini15} as they represent the low-mass, low-luminosity, and probably young-age tail of the FSRQs distribution \citep{berton16, berton17}.

\begin{figure*}[!ht]
    \centering
    \includegraphics[width=0.99\textwidth]{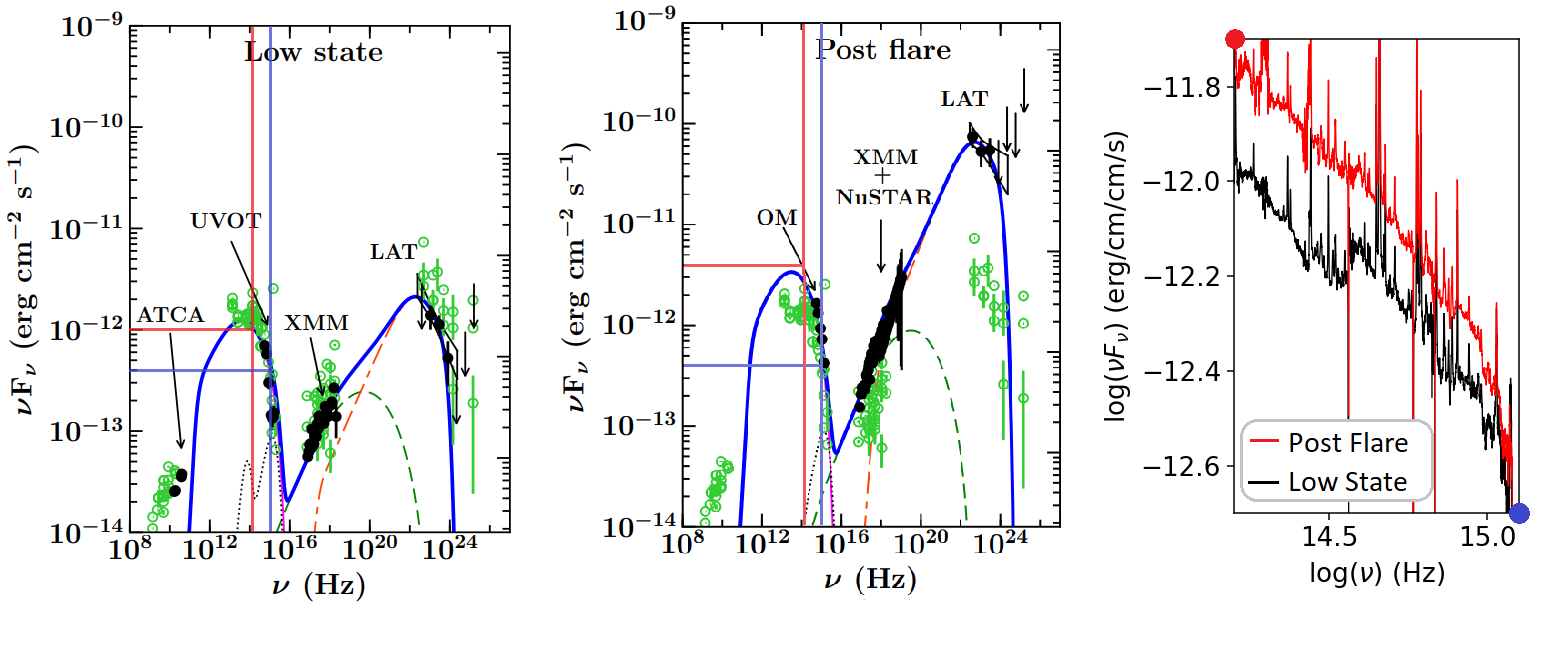}
    \caption{Left and Middle Panel adapted from Figure 6 of \cite{gokus21}. The thin solid pink line shows the synchrotron emission. Dashed green and dash-dash-dot orange lines correspond to the synchrotron self Compton and external Compton processes, respectively. The dotted black line shows the thermal emission from the accretion disk and dusty torus. Right Panel : SP0 as low state in black, and SP2 in red as post-flare, plotted with the same axes as the other panels. The red and blue horizontal and vertical lines in the left and middle panel indicates the y- and x-axis limit plotted in the right panel. The match between the SED and spectrum for both states are very close.}
    \label{fig:sed}
\end{figure*}

PKS 2004-447 (R.A. 20h 07m 55s, Dec. -44d 34m 44s, z$\sim$0.24), the main focus of this paper, is a $\gamma$-ray NLS1. The source has the spectral energy distribution of beam jetted AGN \citep{gallo06, kreikenbohm16, gokus21} and is hosted in a face-on, pseudo-bulge spiral galaxy \citep{kotilainen16}. The radio morphology of PKS 2004-447 suggests an angle relative to line-of-sight of $<50\deg$ \citep{schulz16}, similar to the $\gamma$-NLS1, 3C 286 \citep{berton17, an17, yao21}, although different to other sources with $<10\deg$ angles \citep[e.g.,][]{d13, lister16}. It is less variable in X-rays and radio when compared to other $\gamma$-NLS1s \citep[e.g., see PMN J0948+0022,][]{foschini12}. Most importantly, the source underwent a $\gamma$-ray flare that was detected by the \textit{Fermi}-Large Area Telescope on 2019-10-25, which represents an ideal laboratory to study possible effects of the jet over the BLR.

In this work, we present detailed broad emission lines analysis of X-Shooter data of PKS 2004-447 surrounding the $\gamma$-ray flare. There are two `low-state' spectra taken on 2017-05-28 (SP0) and 2017-06-19 (SP1, both in programme 099.B-0785). After the flare, a `post-flare' spectrum was obtained on 2019-11-25 (SP2), then `reverted-state' spectrum on 2021-05-05 (SP3, both in Director Discretionary Time programme 104.20UC). Note that we deliberately observed the post-flare spectrum, SP2, when the $\gamma$-ray flare reaches the BLR \citep[see][for size estimations]{berton21}. We show, for the very first time, a temporal relation between a $\gamma$-ray event, jet flaring, and BLR changes localised to the redshifted part of the emitting region. This work adopts a standard $\Lambda$CDM cosmology with $H_0 = 70$ km s$^{-1}$ Mpc$^{-1}$ and $\Omega_\lambda=0.7$. All magnitudes are provided in the AB system. 

\section{X-Shooter Observations and Data Reductions}
X-Shooter observes simultaneously in the UVB, VIS, and NIR arms, spanning a total wavelength range of 3,000-25,000 \AA. All four spectra were taken with a 1\arcsec slit. Details of the four X-Shooter spectra are listed in Table~\ref{tab:xshoot-obs}, and they are displayed over the full X-Shooter range in the Appendix (see Figure~\ref{appdx:cts})

For the reduction of raw data, we used the default X-Shooter pipeline provided by ESO\footnote{\url{https://www.eso.org/sci/software/pipelines/index.html\#reflex\_workflows}} with bias, flat, lamp corrections, and response correction with a standard star. We also used ESO Molecfit\footnote{Molecfit Pipeline Team, MOLECFIT Pipeline User Manual, 2021. ESO VLT-MAN-ESO-19550-5772} to remove the telluric absorptions \citep{smette15, kausch15}. SP1 had sufficient signal in the NIR arm of X-Shooter to accurately recover the features and the continuum that was absorbed. We de-reddened the spectra using a \cite{fitzpatrick99} law with E(B-V)=0.0846 \citep{schlafly11} and R$_v$=3.1. We measured the \ion{Na}{i} doublet to obtain the redshift of the host galaxy, obtaining a value of $z=0.24035\pm0.00009$ for SP0, SP1 and SP3, and $z=0.24050\pm0.00009$ for SP2 due to the motion of the Earth. We assume the lower value redshift for calculations in this work, and an operating rest-frame wavelength of 2,418-20,155 \AA.

The airmasses and seeing vary for the four observations, suggesting a need to fine-tune flux calibration. This is usually done by assuming non-varying forbidden-lines, but the different slit orientations sample different parts of the narrow-line region. This variation only affects the forbidden-line, as the continuum and permitted-lines from the central region are well within the 1\arcsec slit. Thus, flux per spectrum is fine-tuned based on the acquisition images (details in Appendix~\ref{appdx:acqui}). The calibration is accurate as the spectra matches the SED from \cite{gokus21} in Figure~\ref{fig:sed}. It is not possible to compare variation in the forbidden-lines across these four spectra due to this systematic.

By averaging SP0, SP1 and SP3, we construct SP013, a higher S/N stacked spectrum representing the long-term average of PKS 2004-447. This smooths out stochastic and systematic variation that is not characterised, and serves as the baseline for comparing with SP2. Excess in SP2$-$SP013 can be easily attributed to the $\gamma$-ray flare.

\begin{table*}
\caption{List of X-Shooter observations.}
\centering
\resizebox{0.9\textwidth}{!}{%
\begin{tabular}{cccccccccc}
\hline
Name & Epoch & Epoch & Airmass & Airmass & Seeing & Exposure & Rotation & S/N & SDSS $r$\\
& (Date) & (MJD) & Start & End &  & time (s) & ($^\circ$) & & (AB mag) \\\hline
SP0 & 2017-05-28 & 57901.206 & 1.453 & 1.231 & 1.05 & 235 & 0 & 13 & 18.448\\
SP1 & 2017-06-19 & 57923.214 & 1.156 & 1.080 & 1.21 & 235 & 40 & 13 & 18.457 \\
SP2 & 2019-11-25 & 58812.027 & 1.636 & 2.106 & 1.27 & 750 & 10 & 24 & 18.188\\
SP3 & 2021-05-05 & 59339.368 & 1.096 & 1.065 & 0.94 & 750 & 50 & 17 & 18.528\\ \hline
\end{tabular}%
}
\tablefoot{Columns: (1) Spectrum name; (2) Observation date; (3) Observation date in modified Julian date (MJD); (4) Airmass at the beginning of the observation; (5) Airmass at the end of the observation; (6) DIMM Seeing during the observation; (7) Exposure time in seconds; (8) Slit rotation as measured with respect to the first spectrum, SP0; (9) Signal to noise ratio measured from the featureless region between 5300-5400\AA; (10) SDSS magnitudes obtained from the flux calibrated acquisition images. }
\label{tab:xshoot-obs}
\end{table*}

\begin{figure*}
    \centering
    \includegraphics[width=0.99\textwidth]{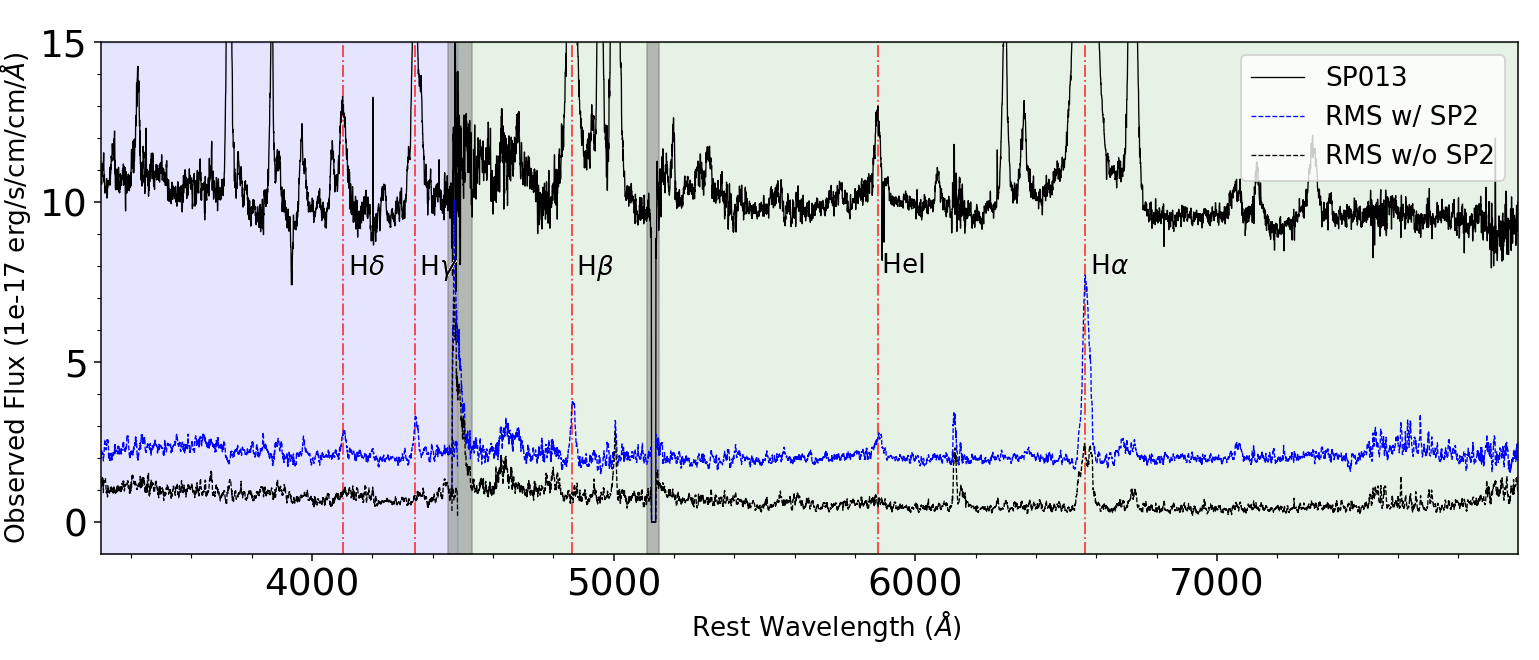}
    \includegraphics[width=0.99\textwidth]{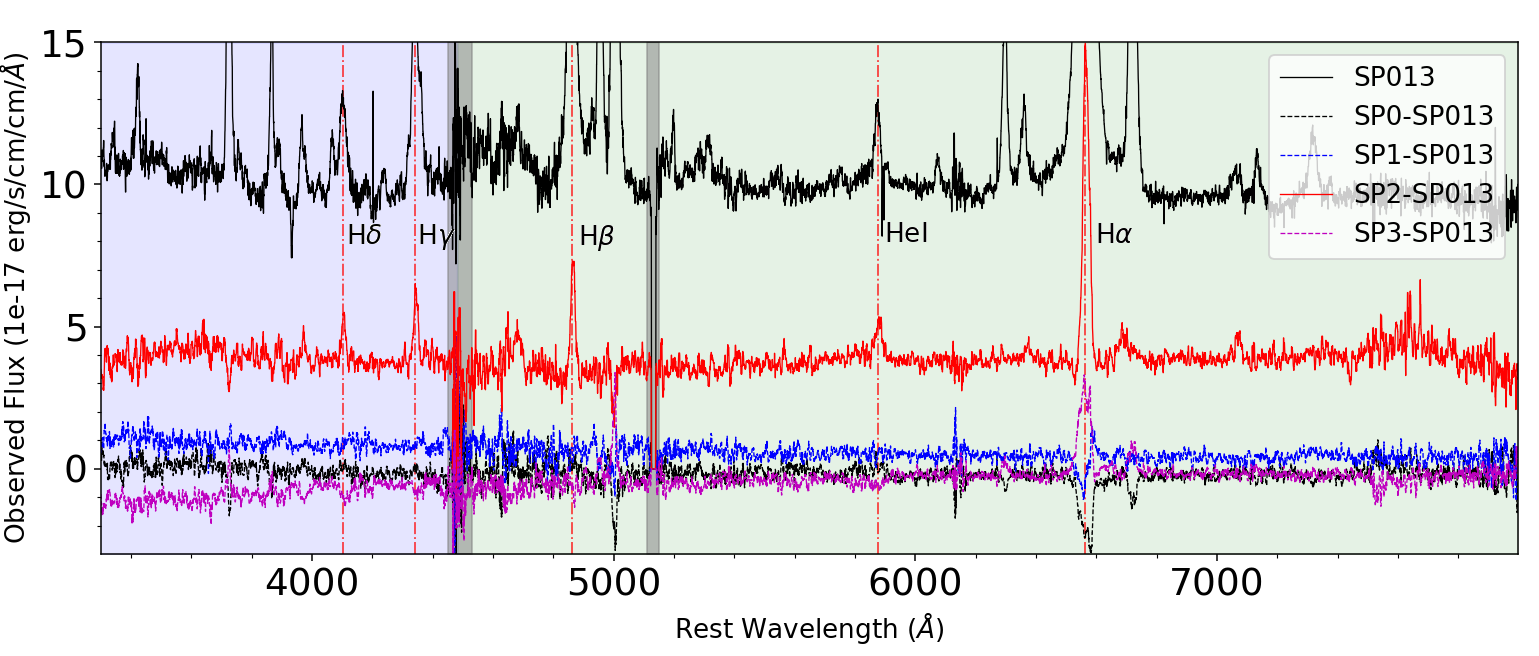}
    \caption{Both panels showing the present of broad line enhancements in the post-flare SP2 spectrum. All plotted lines are indicated by the respective legends, while vertical dash-dotted line indicates the key Balmer and \ion{He}{i}$\lambda$5875 emission line. Gray regions represents sections of the X-Shooter spectra with bad pixels. SP013 in both panels are smoothed with a median filter with a 11 pixel window, while the other lines are smoothed with a 31 pixel window.}
    \label{fig:diffs}
\end{figure*}

\section{Results}
The continuum of PKS 2004-447 is blended. There is; the power-law from the AGN accretion disk; \ion{Fe}{ii} multiplet emission; host galaxy that is resolved in the aperture \citep{kreikenbohm16}; synchrotron emission from the jet. Its UV-NIR frequencies are dominated by synchrotron emission, resulting in a negative slope instead of the usual positive power-law \citep{gokus21}. During the flare, the flux is boosted across all X-Shooter wavelengths, with greater increase for lower frequencies.

To characterise the differences between the four datasets, we construct RMS spectra using \cite{peterson04} Equation 3. Figure~\ref{fig:diffs} shows the RMS spectra with and without SP2. Without SP2, no features appear above the noise around H$\delta$, H$\gamma$, H$\beta$ and \ion{He}{i}$\lambda$5875. With SP2, those lines show noticeable enhancements. We investigate each spectrum's variability relative to SP013 and find that the enhancement only appears in SP2, suggesting that the broad emission lines in the Balmer and helium series changed temporarily in SP2, likely related to the $\gamma$-ray flare.

We can also see the effects of mismatched slit orientations here. While SP1$-$SP013 has somewhat clean subtractions in the forbidden lines, SP0$-$SP013 and SP3$-$SP013 do not. This makes interpreting the changes in H$\alpha$ difficult. While this emission line is also affected, it is hard to disentangle observed and systematic variation on the [\ion{N}{ii}] line

\begin{figure*}
    \centering
    \includegraphics[width=0.98\textwidth]{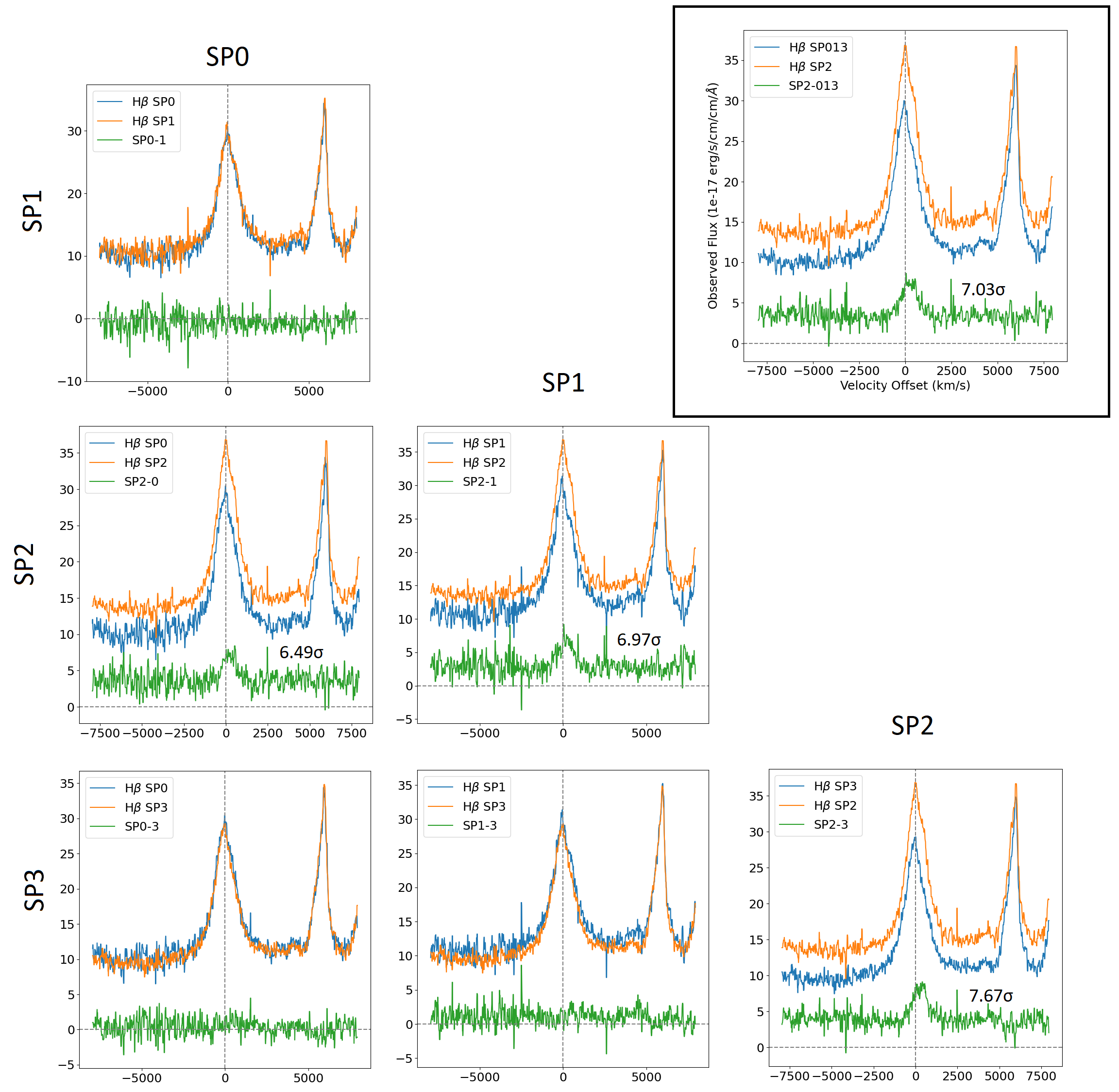}
    \caption{A grid of plots showing the H$\beta$ region from the four X-SHOOTER spectra. All lines are smoothed with a median filter with a 5 pixel window, which is much much narrower than any features that could be present. In each panel, two spectra (blue and orange line) and their subtracted result (green line) are plotted with observed flux in y-axis and Velocity offset in x-axis. The two spectra in each panel depends on the row and column as indicated in large font. For example, the bottom right corner is plotting SP2 and SP3. In the top left corner, the panel with a black box is plotting SP013 and SP2. We also indicate the 0 value in the y- and x- axes for every panel with vertical and horizontal dashed lines. These visual aids emphasise the non-zero redshift in the emission line excess in panels involving SP2. The statistical significance of the excess is also given in the relevant panels, see footnote 3 for details. We see that in all other panels without SP2, the green line is consistent with 0, indicating almost no variation outside of the flare.}
    \label{fig:excess}
\end{figure*}

A closer look at the enhancement in H$\beta$ as well as comparison of all epochs is in Figure~\ref{fig:excess}. Outside of SP2, the H$\beta$ line remains constant with the subtracted spectrum, consistent with a $\sim0$ mean value within the noise level. All of the H$\beta$ enhancement is significant\footnote{Calculated with the peak flux value of the enhancement and the RMS of the SP2 continuum. The window for RMS is 5300-5400\AA\ the same as used in S/N calculation in Table~\ref{tab:xshoot-obs}} at $>6\sigma$. We also discover that the peak of the H$\beta$ excess is redshifted by\footnote{error quoted here accounts for line fitting uncertainty and X-Shooter resolution of 35km s$^{-1}$ at H$\beta$} 240$\pm$50km s$^{-1}$ (4.8$\sigma$), consistent across all the subtracted spectrum. We denote this enhancement with a redward bias the 'red-excess'. All of the Balmer lines and \ion{He}{i}$\lambda10830$ exhibit the red excess, but it is less obvious in \ion{He}{i}$\lambda5875$ (see Appendix Figure~\ref{appdx:excess}). Weaker enhancements are seen in Pa$\beta$ and Pa$\alpha$, however they are not significantly redshifted. \ion{O}{i} and \ion{Mg}{ii} have no enhancements.

Line fitting is not trivial for PKS 2004-447 as the continuum is not a power-law. Approximations work for individual epochs, but will be consistent across all four spectra. This is because both the emission lines and continuum can vary, so it is not possible to disentangle the source of any measured variation. Therefore, we only fit SP013 to provide average BLR properties and SP2$-$SP013 to estimate red-excess properties (details in Appendix B). Enhancements and red-excess are present independently and not due to continuum modelling.

The width of all lines are around FWHM$=$1600 km s$^{-1}$. All Balmer and Paschen lines have $<50$ km s$^{-1}$ velocity offset while all \ion{He}{i} lines are blueshifted to 85 km s$^{-1}$. \ion{O}{i}$\lambda$8446 is redshifted to 180 km s$^{-1}$, suggesting an in-flow. 

For the red-excess, we measure a FWHM$\sim$1000 km s$^{-1}$ for all Balmer and Paschen lines, but the offset differs for different line series. H$\delta$, H$\gamma$ and H$\beta$ are redshifted by $\sim$250 km s$^{-1}$, H$\alpha$ is redshifted by 180 km s$^{-1}$. \ion{He}{i} red-excess are comparatively weaker for this analysis, but we observe roughly the same width and up to 340 km s$^{-1}$ of redshift. As mentioned, Paschen lines do not have a noticeable redshift, suggesting that the red-excess is more prominent towards the blue.

\section{Discussion and Conclusion}
\subsection{Red-excess from observational effects?}
We explore the possibility that the red-excess is caused by observational effects. This is a possibility since the observing conditions between all spectra are different. Namely, SP2 that is the spectrum closest to the flare and the spectrum that is showing the red-excess, is observed under the worse conditions relative to the rest (see Table~\ref{tab:xshoot-obs}).

The immediate cause of concern with SP2 is flux dimming caused by high ending airmass at $\sim$2.1 and seeing 1.2\arcsec. The drop in flux caused by poor seeing or field differential dispersion is uniform, and this is corrected when we calibrate the flux photometrically (see Appendix~\ref{appdx:acqui}). Also, as seen in Table 1, SP0 and SP1 have similar S/N given similar exposure times, followed by SP3 and SP2 with the best S/N since PKS2004-447 is at its brightest. This demonstrates that the quality of spectra is more dependant on exposure times and source brightness. 

Moreover, PKS 2004-447 has a scale of 3.8 kpc/\arcsec at z=0.24. For a BLR radius of 15 light days (Berton et al. 2021), this corresponds to a diameter of 2.5$\times$10$^{-5}$ kpc, which covers $\ll$1\arcsec. With a slit of 1\arcsec, this BLR scale is too small for airmass and seeing to have any relevant effect to the emission lines. Even if it did, it should be causing the BLR to be dimmer, not brighter. It would also be affecting the NLR and \ion{Mg}{ii} line, which is not observed. 

Finally, VLT/XShooter is equipped with Atmospheric Differential Correctors (ADCs) that prevent spectral distortions for airmass lower than 2.5. Therefore it is unlikely for the red-excess to be an observational artefact.

\subsection{BLR Geometry}
The difference in peak offset between H (close to systematic redshift) and He (almost 100 km s$^{-1}$ blueshifted) indicates a stratified BLR, while the larger FWHM in \ion{He}{i} indicates that He is at a smaller radial distance. This is consistent with the theory \citep{korista04} and observations from velocity-resolved reverberation mapping \citep[e.g.][]{kollatschny03, bentz10}. However the large uncertainty associated with \ion{He}{i} (see Table~\ref{tab:lines}) makes it impossible to meaningfully conclude the BLR geometry based on this information. 

The kinematics of the Balmer lines are well measured and so is the velocity offset of the redshifted \ion{O}{i} line. The latter implies the presence of an in-flow. Considering the source is a jetted AGN, the viewing angle is likely face on with a $<50\deg$ constrain from \cite{schulz16}. Under a disk-wind model, for Balmer lines to not have a significant peak offset implies the emitting region is moving perpendicular to the line of sight. Building on this assumption, two scenarios exist for placing the other emitting regions. The first is a single stream of material that follows the locally optimally emitting clouds \citep[LOC,][]{baldwin95} model, with \ion{He}{i}, Paschen and Balmer lines, ending with \ion{O}{i}. This implies a greater push at the start of the stream, thus the blueshifted peaks of \ion{He}{i}. The second is a dual-stream disk-model \citep[e.g.,][]{yong17} with He wind radially closer and at higher inclination than the H wind, not placing Oi in-flow at any required position. Reverberation mapping will be needed to fully investigate the structure, and a higher resolution spectrum with better S/N focusing on the Paschen, \ion{He}{i}, \ion{He}{ii} lines is required.

Our results reveal perplexing line fluxes and ratios. The Balmer decrement is very steep at H$\alpha$:H$\beta$:H$\gamma$:H$\delta$ = 4.67:1.00:0.31:0.15. Also, Pa$\alpha$/Pa$\beta$=1.08, but Pa$\alpha$/H$\beta$=0.50 is very low (all values are derived from fluxes presented in Table~\ref{tab:lines}). A reddening effect does not explain this, as a consistent range of case B values for H$\gamma$/H$\beta$ would worsen the Pa$\alpha$/H$\beta$ ratio and result in H$\alpha$/H$\beta$ being too low. A differential in optical depth could explain the Balmer ratios, yet this would require an unusually high optical depth of $\gg$1000\citep[see fig. 9 of][]{davidson79}. Balmer line ratios are known to be correlated with velocity offset, where the H$\alpha$/H$\beta$ increases towards the peak of a line profile \cite{stirpe91}. This could be a feasible explanation as the line profile observed in PKS 2004-447 is highly dominated by the peak. Alternatively, the emission lines may not be completely recombination driven, and a partial local thermodynamical equilibrium exist and the Saha-Boltzmann relations holds for part of the BLR \citep{popovic03}. This is supported by the Balmer ratios in PKS 2004-447 being similar to 3C 390.3 and 3C 382 from that work. 

\subsection{Understanding of red-excess}
The red-excess and the flare are correlated events. This is the first observed event where a $\gamma$-ray flare is associated to a local change in the BLR with distinct kinematic properties. If multiple spectra were taken as soon as the flare started, we could assess if the red-excess appeared before or after the flare. If before, we would observe it fading, implying a link to the cause of flare. If after, we would see it growing, suggesting a direct contribution of the jet to the red-excess.

The most curious aspect of the red-excess is that it is redshifted. Here we speculate about four possible scenarios on how it is formed. 

(1) The simplest scenario is a bipolar outflow, with receding gas visible and approaching gas partially obscured. This is motivated by the red-excess being more prominent towards the blue and absent for the IR Paschen lines. The \ion{O}{i} in-flow also supports this model. The flare would reverberate throughout both flows equally, reaching most of the material within a month. However, shorter wavelength emissions from the approaching ions would be obscured, biasing a redward enhancement.

(2) The $\gamma$-ray flare directly creates the red-excess. If the event occurred above the H and He emitting region, a shock-wave from the event would push the gas down, introducing more material into the emitting region. In this scenario, we should also observe an red-excess in \ion{Mg}{ii} and \ion{O}{i}, but that is not seen. This is a point of contention for this scenario. Using the BLR radius calculated by \cite{berton21} of $\sim$15 light days, a FWHM$\sim$1500 km s$^{-1}$, and assuming $t_{dyn}\approx R_{BLR}/$FWHM gives a dynamical timescale of 8 years. This is the time needed for the introduced material to re-equilibrate, which is longer than the 1.5 years between SP2 and SP3. Reconciling this requires that the event occurred at a distance $<$3 light days, which is consistent with the predicted location of $\gamma$-ray production at $\sim10^3 r_g \sim$1 light day \citep{ghisellini10}. 

(3) The flaring event led to an increased illumination onto a region close enough to the black hole to experience strong gravitational redshift. A shift of 250 km s$^{-1}$ yields an emitting radius of $\sim$1 light day (with M$_{bh}=1.5\times10^7$M$_{\odot}$ from \citealp{berton21}), which is consistent with an increased illumination from the $\gamma$-ray flare itself. The emergence of the red-excess should be instant due to its proximity to the black hole and $\gamma$-ray source. However, its visibility a month after the flare poses an issue, even accounting for gravitational time dilation.

(4) The accretion disk or BLR undergo a structural change that leads to the flare and increased illumination in a specific part of the BLR. This region could be close or far from the black hole. If close, the duration of the illumination is not tied to the short-lived flare unlike in scenario (3). If far, since \ion{O}{i} is an established in-flow, it might imply the outermost parts of the BLR are always in-falling. This region might contribute to part of the red-wing observed in the permitted line profiles, which can be verified with velocity resolved reverberation mapping. The relevant timescale for this interaction is not well understood, as the in-falling timescale based on the viscosity parameter has been shown to be flawed as demonstrated by Changing-Look AGN \citep{lawrence18}.

All of the speculated scenarios would be extremely interesting to study, and we encourage frequent observation immediately after a flare for all $\gamma$-NLS1s.

\begin{acknowledgements}
We appreciate the effort and time of the anonymous referee in improving on this manuscript. W.-J.H., M.B., and E.S. acknowledge the support of the ESO studentship programme. 
M.B. and E.C. are ESO fellows. 
E.C. acknowledges support from ANID project Basal AFB-170002.
Based on observations collected at the European Southern Observatory under ESO programmes 099.B-0785 and 104.20UC.
ARL acknowledges the support by FONDECYT Postdoctorado project No. 3210157.
\end{acknowledgements}

\bibliographystyle{aa}
\bibliography{bib2}

\begin{appendix}
\section{Flux Calibration from Acquisition Images}\label{appdx:acqui}

\begin{figure}
    \centering
    \includegraphics[width=0.47\textwidth]{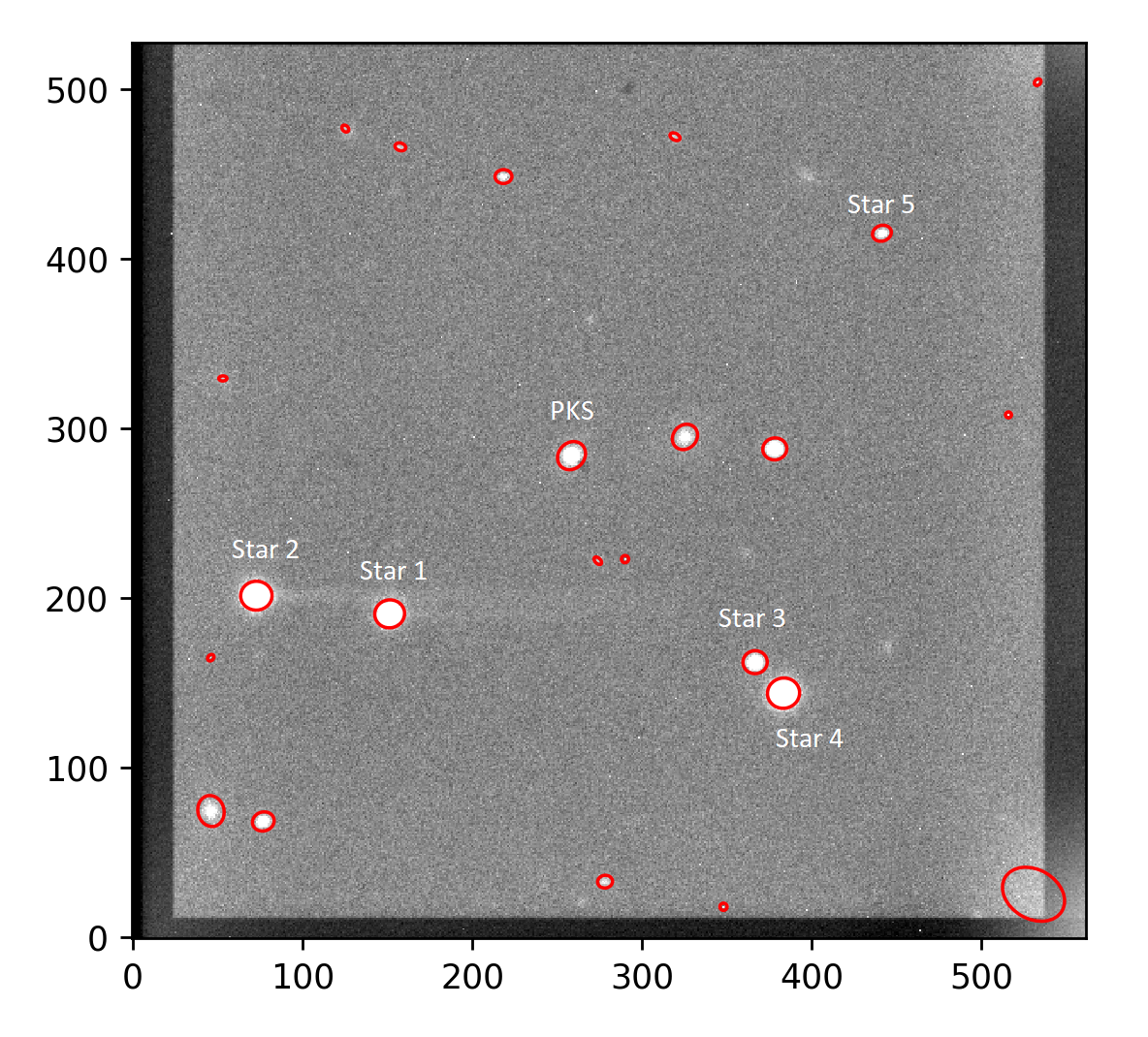}
    \caption{Acquisition image from SP0 with the sources selected by {\sc SEP} marked in red ellipses. While there are a large number of false selections, we are only concerned with the ones that are labelled.}
    \label{fig:sep-acqui-photo}
\end{figure}

The final flux calibration of the spectra utilises the raw acquisition images taken under the $r$-band Sloan filter \citep{lenz98}. We calibrate these images by tweaking the zero point such that stars in the field (see Figure~\ref{fig:sep-acqui-photo}), identified using Gaia's proper motion data \citep[Gaia EDR3 data was used]{brown21}, have a constant magnitude across all observations. The baseline magnitude is taken from Pan-STARRS \citep{tonry12} that operates with a similar filter set. To measure star brightness, we use the python package {\sc SEP} \citep{barbary16}, a python version of Source Extractor \citep{bertin96}. We focus on PKS 2004-447 and the five stars that are labelled. The fluxes are then counted from 2.5$\times$ the kron radius, resulting in the {\sc MAG\_AUTO} of Source Extractor. The kron radius is the first moment of the surface brightness light profile and sufficient amount of the total flux is contain within 2.5$\times$ this radius \citep{kron80}.

The zero points of each epoch were tweaked such that `Star 1' remains almost constant at 16.650 magnitude. The other stars are then only variable on the order of 0.01 magnitude. `Star 5' being much dimmer is an exception with variability on the order of 0.1 magnitude. This choice of zero points implies that PKS2004-447 has minimal variability between SP0 and SP1, is at the brightest at SP2 and dimmest at SP3 for a maximum variation of 0.34 magnitude. 

\begin{figure*}
    \centering
    \includegraphics[width=0.99\textwidth]{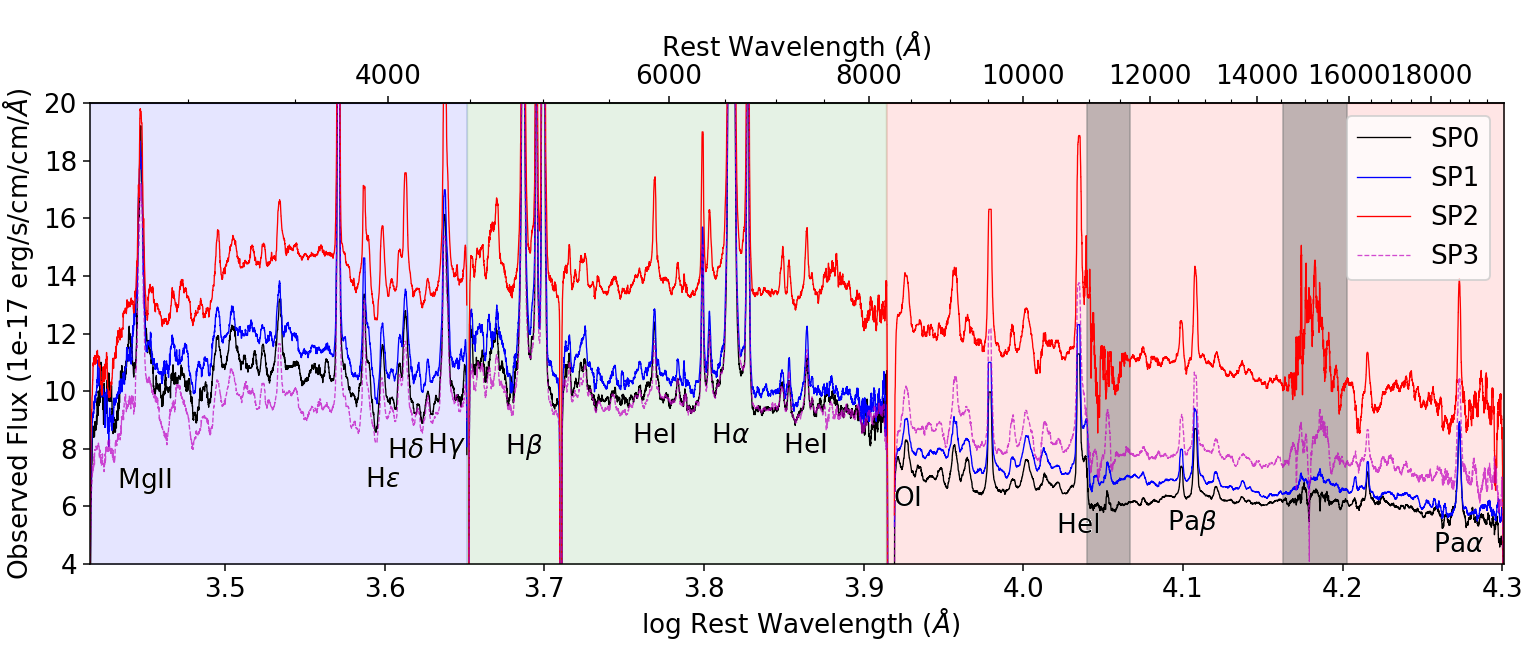}
    \caption{Full rest-frame wavelength range of SP0, SP1, SP2 and SP3. The colour sections indicates the three arms, while the two grey sections in the NIR arm marks the most problematic telluric regions that Molecfit was not able to recover.}
    \label{appdx:cts}
\end{figure*}

Assuming AB magnitude, we converting PKS 2004-447 magnitudes into average fluxes across SDSS R-band with information from the Spanish Virtual Observatory Filter Profile Service \citep{rodrigo20}. Finally, we scale the spectra of each epoch to match this derived flux level. Note that this process has converted slit continuum flux to kron aperture continuum flux to allow comparison or variability between epochs without worrying about airmass losses and calibration uncertainties. This results in the emission lines fluxes to be accurate relatively across epochs, but over-estimated by themselves with an average factor of 1.61. By undoing the scaling we will recover the line luminosities when required. Also, slit differences between narrow emission lines are still not accounted for. The full X-Shooter view of this source is shown in Figure~\ref{appdx:cts}

\begin{table}
\caption{ZP in the table refers to zero point. Errors are not given as they are $<0.1\%$ from Source Extractor, but the larger uncertainty comes from variations between epochs.}
\centering
\resizebox{0.45\textwidth}{!}{%
\begin{tabular}{c|cccc}
\hline
Object & SP0 & SP1 & SP2 & SP3 \\ \hline
ZP & 29.015 & 29.077 & 29.75 & 29.76 \\
PKS & 18.448 & 18.457 & 18.188 & 18.528 \\
Star 1 & 16.650 & 16.650 & 16.649 & 16.653 \\
Star 2 & 16.005 & 16.001 & 16.013 & 16.008 \\
Star 3 & 15.940 & 15.925 & 15.940 & 15.921 \\
Star 4 & 18.667 & 18.626 & 18.664 & 18.679 \\
Star 5 & 20.203 & 20.114 & 20.129 & 20.029 \\ \hline
\end{tabular}%
}
\tablefoot{Columns: (1) Object label; (2-5) SDSS R-band magnitudes, with zero points defined in table, of the objects in acquisition images associated to SP0, SP1, SP2, and SP3 respectively.}
\label{tab:acqui-phot}
\end{table}

\section{Spectral Analysis and Line-Fitting}\label{appdx:fitting}
We used the python script PyQSOFit by \cite{pyqsofit} that utilises the least squares fitting package {\sc kmpfit}. The residuals are weighted based on the variance spectra. PyQSOFit has the functionality to fit continuum and lines. For continuum fitting, while optimising fitting range to feasibility, we approximate the rest-frame range from 3300-7500\AA. This encompasses half of the UVB and the entire VIS arm. We fit with a free power-law, a third-order polynomial, and the \ion{Fe}{ii} template from \cite{boroson92}. The NIR arm is fitted separately and can be approximated by a power-law in its entirely. 

For line-fitting, we used a modified version of PyQSOFit called PyQSOFit\_SBL\footnote{\url{https://github.com/JackHon55/PyQSOFit\_SBL}}. This version allows for skewed Gaussian and Voigt profile fitting. It also has the functionality to link the parameters of an unlimited amount of fitted components. This reduces the degree of arbitrariness as we de-blend the H$\alpha$ line. It also reduces the number of individual parameters to vary in a fitting, as well as increasing the sampling of the parameter of each related component for a more accurate result. While the script is able to perform bootstrapping for error estimation, the more realistic errors would be from the variance of each measured value throughout the three spectra for the SP013.

\begin{figure*}
    \includegraphics[width=0.33\textwidth]{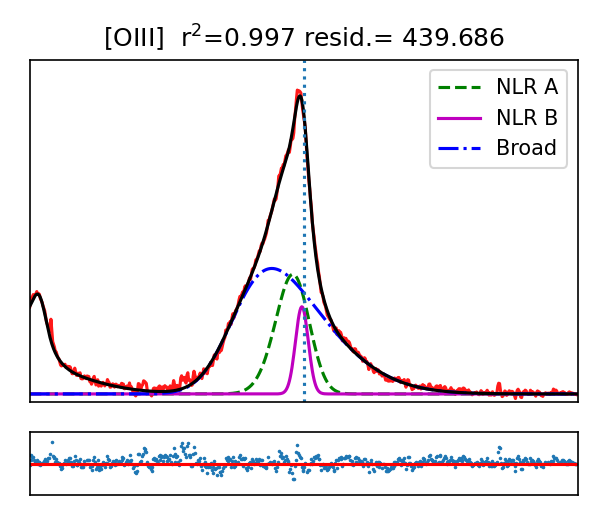}
    \includegraphics[width=0.33\textwidth]{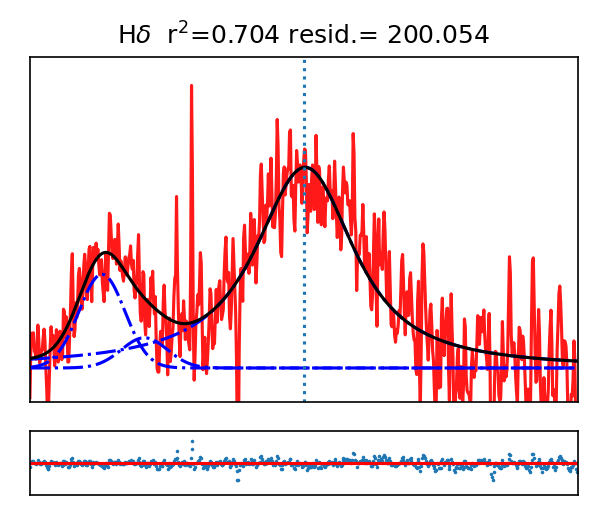}
    \includegraphics[width=0.33\textwidth]{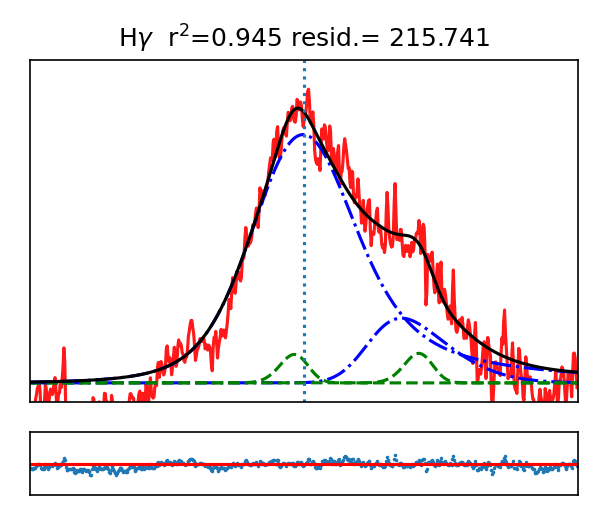}
    \includegraphics[width=0.33\textwidth]{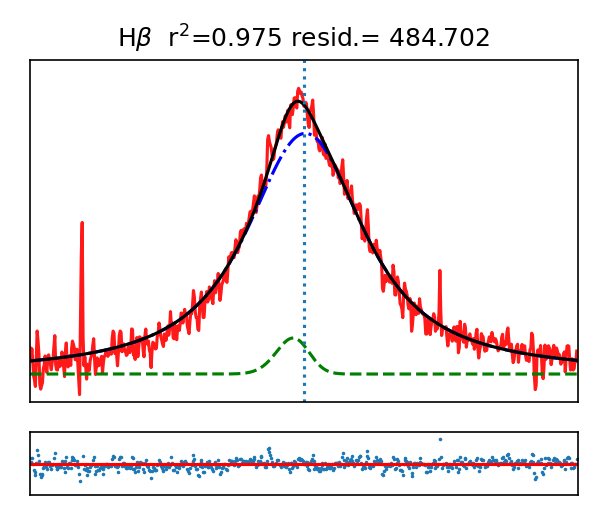}
    \includegraphics[width=0.33\textwidth]{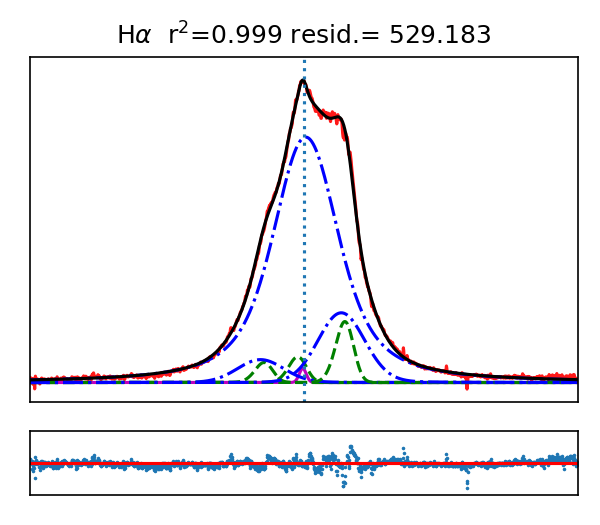}
    \includegraphics[width=0.33\textwidth]{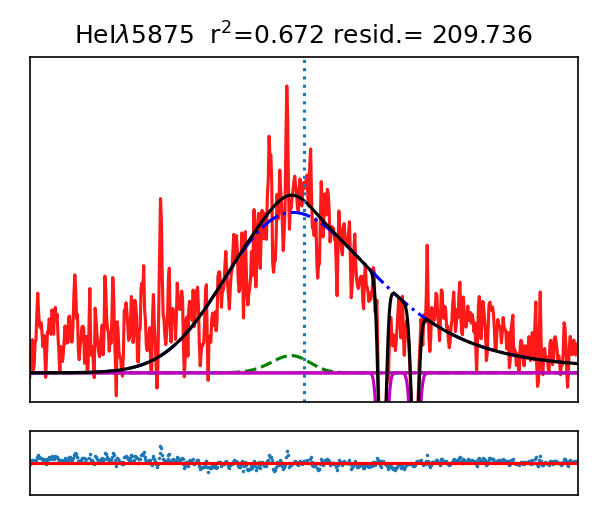}
    \includegraphics[width=0.33\textwidth]{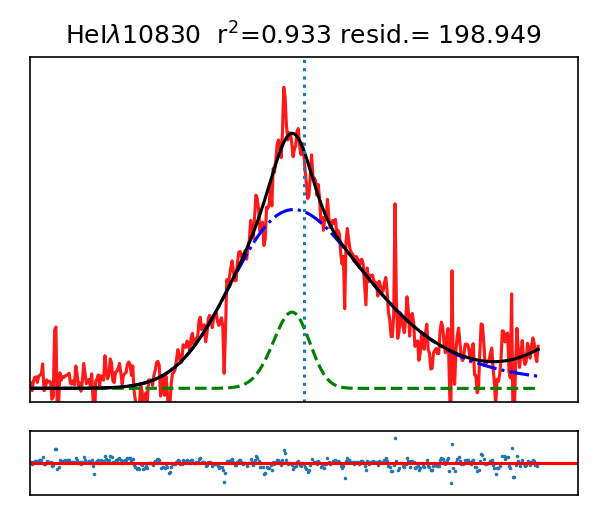}
    \includegraphics[width=0.33\textwidth]{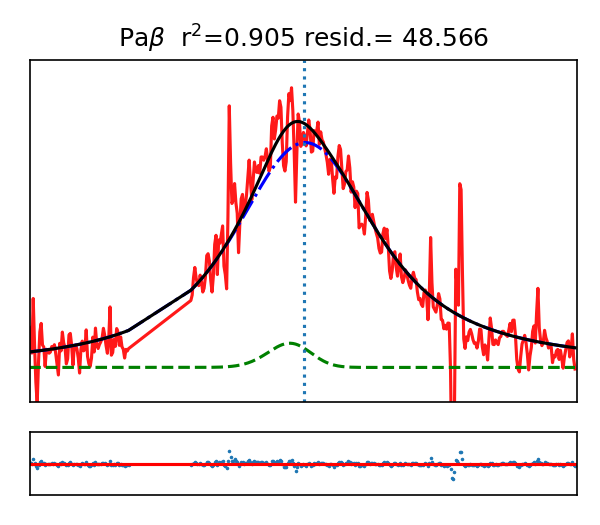}
    \includegraphics[width=0.33\textwidth]{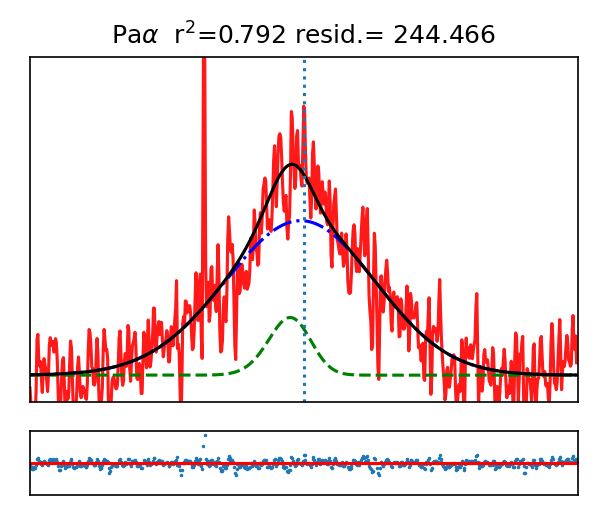}
    \includegraphics[width=0.33\textwidth]{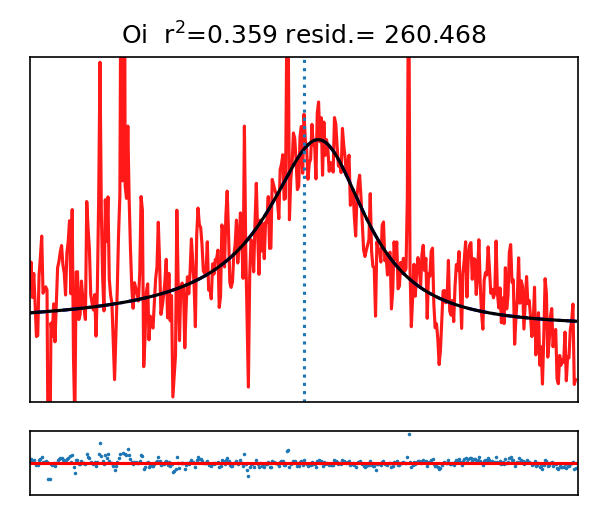}
    \includegraphics[width=0.33\textwidth]{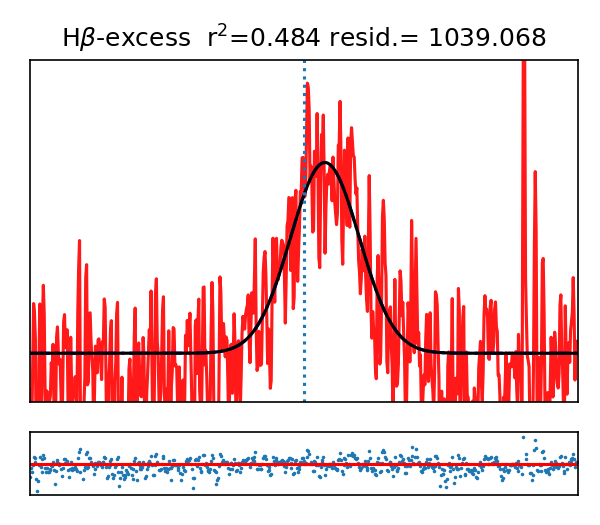}
    \caption{Line fitting on continuum subtracted SP013. Each panel has a top plot to show the components used to fit, and a bottom plot to show the residuals. All lines are colour coded based on the legend of [\ion{O}{iii}]. The last plot is showing the red-excess fitting for H$\beta$, which has the best and isolated signal. It is not conclusive if the feature can be approximated with a Gaussian.}
    \label{fig:fitting}
\end{figure*}

One key finding through trial and error is that the forbidden lines need to be fitted with three components. These are, a broad skewed Voigt for the outflows, a narrow skewed Gaussian intermediate component (NLR A), and a very narrow Gaussian component with very little offset relative to the host galaxy (NLR B). The NLR A and NLR B components are defined from [\ion{O}{iii}]$\lambda$5007, however the broad component is different for every line. Either coincidentally, or physically, the narrow components of the permitted lines can be associated to NLR A and NLR B, so these are used to simplify the fitting procedure. The line profiles will be described here and are also shown in Figure~\ref{fig:fitting} including [\ion{O}{iii}], and the results of the fitting is summarised in Table~\ref{tab:lines}.

\textbf{Balmer Lines} - In the SP013 spectrum, the broad component of H$\delta$, H$\gamma$ and H$\beta$ are similar. H$\gamma$ and H$\beta$ has a detectable NLR A component. H$\gamma$ fitting is considered problematic as it is heavily blended with [\ion{O}{iii}]$\lambda$4363 and the continuum approximation over-subtracts the blue-wing. 

H$\alpha$ is complicated as it is directly blended with [\ion{N}{ii}], and potentially blended with \ion{He}{i}$\lambda$6678 and [\ion{S}{ii}] on the red-wing. The final fitting to measure H$\alpha$ broad emission lines uses a total of seven components; four for the [\ion{N}{ii}] doublet with two NLR A and two broad components set to the theoretical ratio; three for H$\alpha$ with NLR A, NLR B and a broad component that is associated to H$\beta$ (no fixed, but has to be similar). The blending from \ion{He}{i}$\lambda$6678 and [\ion{S}{ii}] was not significant, and the final component of [\ion{N}{ii}] was comparatively too weak to alter the H$\alpha$ broad line. 

\textbf{Paschen Lines} - Paschen lines observed ranges from Pa$\epsilon$ to Pa$\alpha$, however we are only able to acquire reliable line-fitting measurements for Pa$\beta$ and Pa$\alpha$. Pa$\epsilon$ is heavily blended with [\ion{S}{iii}]$\lambda$9531, Pa$\delta$ and Pa$\gamma$ are both dominated by noise that is introduced by telluric absorption. We find a broad component and a NLR A component for both Pa$\beta$ and Pa$\alpha$. Broad Pa$\alpha$ is problematic due to telluric noise, causing unreliable continuum definition. 

\textbf{Helium Lines} - Of the observed He lines, we could only reliably measure \ion{He}{i}$\lambda$10830. The only \ion{He}{ii} line is at $\lambda$4686, but sits in the region with heavy \ion{Fe}{ii} emission, which we are unable to constrain. \ion{He}{i} $\lambda$3889 and
$\lambda$7065 are very weak, the former suffers from continuum definition issues and the latter and suffers from profile blending from multiple lines. \ion{He}{i} $\lambda$5876 has continuum definition issues, but we are able to acquire a reliable fit by assuming it having the exact same profile as \ion{He}{i}$\lambda$10830. \ion{He}{i} $\lambda$5876 and $\lambda$10830 both has a significant NLR A component.

\textbf{Red-excess}
All of the red-excess are fitted with a single non-skewed Gaussian component, with no width and offset constrains. The measurement for \ion{He}{i}$\lambda$5876 was extremely problematic as the continuum is not well defined in that region, resulting in contamination from continuum variation. \ion{He}{i}$\lambda$10830 also suffers from contamination from telluric correction variation. As such, their widths and peak offsets are not as reliable as the Balmer lines. We also observed red-excess on \ion{He}{i}$\lambda$6678, but this is heavily contaminated by [\ion{S}{ii}]$\lambda$6725 and H$\alpha$ variations. While not listed because of the inaccurate measurement, it is a significant red-excess with a FWHM$=1765\pm425$ km s$^{-1}$, a redshift of 620$\pm$600 km s$^{-1}$ and a flux of 21.52$\pm8.52\times10^{-17}$ erg/s/cm/cm.

\begin{table*}
\caption{Summary of permitted lines that are observed, that are fitted, and red-excess that are observed and fitted. Undef.Ctm comments are lines that we are not able to fit as we are unable to accurately define the continuum around the line. As for Telluric, Problematic and Blend comments, refer to the text in the Appendix~\ref{appdx:fitting}. Empty entries in the spectral analysis measurements means that the feature was not fitted. Negative values in the offset refers to a blueshift.}

\resizebox{0.97\textwidth}{!}{%
\begin{tabular}{lrl|rrrr|rrr}
\hline
 &  & & \multicolumn{4}{c|}{SP013 Spectrum}  & \multicolumn{3}{c}{Red-excess} \\
\multicolumn{1}{c}{Line} & \multicolumn{1}{c}{Wavelength} & \multicolumn{1}{c|}{Comment} & \multicolumn{1}{c}{FWHM} & \multicolumn{1}{c}{Offset} & \multicolumn{1}{c}{Kurtosis} & \multicolumn{1}{c|}{Flux 10$^{-17}$} & \multicolumn{1}{c}{FWHM} & \multicolumn{1}{c}{Offset} & \multicolumn{1}{c}{Flux 10$^{-17}$} \\
 & \multicolumn{1}{c}{(\AA)} & & \multicolumn{1}{c}{(km/s)} & \multicolumn{1}{c}{(km/s)} &  & \multicolumn{1}{c|}{(erg/s/cm$^2$)} & \multicolumn{1}{c}{(km/s)} & \multicolumn{1}{c}{(km/s)} &  \multicolumn{1}{c}{(erg/s/cm$^2$)} \\
\hline
  \multicolumn{10}{c}{UVB   Continuum Issues} \\
\hline
\ion{Mg}{ii} & 2799.00 & Undef. Ctm &  &  &  &  & &  & \\
\ion{He}{i} & 3888.647 & Undef. Ctm &  &  &  &  & &  & \\
H$\epsilon$ & 3970.079 & Undef. Ctm &  &  &  &  & &  & \\
H$\delta$ & 4101.742 &  & 1580 & 0 & -0.01 & 56.68 & 905 & 240 & 12.50 \\
 &  &  & $\pm$25 & 0 & 0 & $\pm$5.07 & $\pm120$ & $\pm$50 & $\pm$4.73 \\[1mm]
H$\gamma$ & 4340.471 & Problematic & 1580 & 0 & 0.22 & 115.26 & 1030 & 285 & 25.37 \\
 &  &  & $\pm$25 & 0 & $\pm$0.05 & $\pm$2.21 & $\pm30$ & $\pm$35 & $\pm$2.34 \\
\hline
  \multicolumn{10}{c}{VIS} \\
\hline

\ion{He}{ii} & 4685.710 & Undef. Ctm &  &  &  &  & &  & \\
H$\beta$ & 4861.333 & & 1580 & 0 & -0.01 & 375.86 & 905 & 240 & 37.60 \\
 &  &  & $\pm$25 & 0 & 0 & $\pm$7.06 & $\pm80$ & $\pm$40 & $\pm$4.87 \\[2mm]
\ion{He}{i} & 5875.624 & Problematic & 1660 & $-85$ & 1.89 & 58.51 & 795 & 220 & 16.18 \\
 &  &  & $\pm$175 & $\pm$40 & $\pm$0.95 & $\pm$8.48 & $\pm195$ & $\pm$85 & $\pm$3.96 \\[2mm]
H$\alpha$ & 6562.819 & Blend & 1565 & 45 & 0 & 1755.63 & 1130 & 180 & 179.52 \\
 &  &  & $\pm$60 & $\pm$35 & 0 & $\pm$224.10 & $\pm200$ & $\pm$60 & $\pm$65.6 \\[2mm]
\ion{He}{i} & 7065.196 & Problematic &  &  &  &  & 820 & 340 & 9.69 \\
 &  &  &  &  &  &  & $\pm150$ & $\pm$20 & $\pm$1.71 \\
\hline
  \multicolumn{10}{c}{NIR   Few Telluric Issues} \\
\hline

\ion{O}{i} & 8446.359 &  Telluric & 1560 & 180 & 0 & 97.15 & &  & \\
 &  &  & $\pm$225 & $\pm$75 & 0 & $\pm$13.58 & &  & \\
Pa$\epsilon$ & 9545.969 & Blend  &  &  &  &  & &  & \\
Pa$\delta$ & 10049.368 & Telluric  &  &  &  &  & &  & \\
\ion{He}{i} & 10830.340 & & 1660 & $-85$ & 1.89 & 268.68 & 1320 & 150 & 103.06 \\
 &  &  & $\pm$175 & $\pm$40 & $\pm$0.95 & $\pm$40.54 & $\pm$155 & $\pm$110 & $\pm$35.09 \\
Pa$\gamma$ & 10938.086 & Telluric  &  &  &  &  & &  & \\

Pa$\beta$ & 12818.072 &   & 1500 & 0 & 0 & 172.11 & 1100 & 25 & 30.97 \\
 &  &  & $\pm$60 & 0 & 0 & $\pm$25.28 & $\pm$70 & $\pm105$ & $\pm$12.26 \\[2mm]
Pa$\alpha$ & 18750.976 & Telluric  & 1550 & $-25$ & $-0.7$ & 186.14 & 760 & 35 & 37.01\\
 &  &  & $\pm$225 & $\pm$70 & $\pm$0.58 & $\pm$50.19 & $\pm$180 & $\pm$95 & $\pm$20.66 \\
\hline
\end{tabular}%
}
\tablefoot{Columns: (1) Emission line name; (2) Rest-frame wavelengths measured in air in \AA; (3) Specific comments regarding issues in line-fitting the profile; (4) FWHM of the line profile from SP013 in km/s; (5) Velocity offset of the line profile from SP013 in km/s; (6) Kurtosis or skew of the line profile from SP013; (7) Flux of the emission line from SP013 in 10$^{-17}$ erg/s/cm/cm; (8) FWHM of the red-excess in km/s; (9) Velocity offset of the red-excess in km/s; (10) Flux of the red-excess in 10$^{-17}$ erg/s/cm/cm;}
\label{tab:lines}
\end{table*}

\begin{figure*}
    \includegraphics[width=0.95\textwidth]{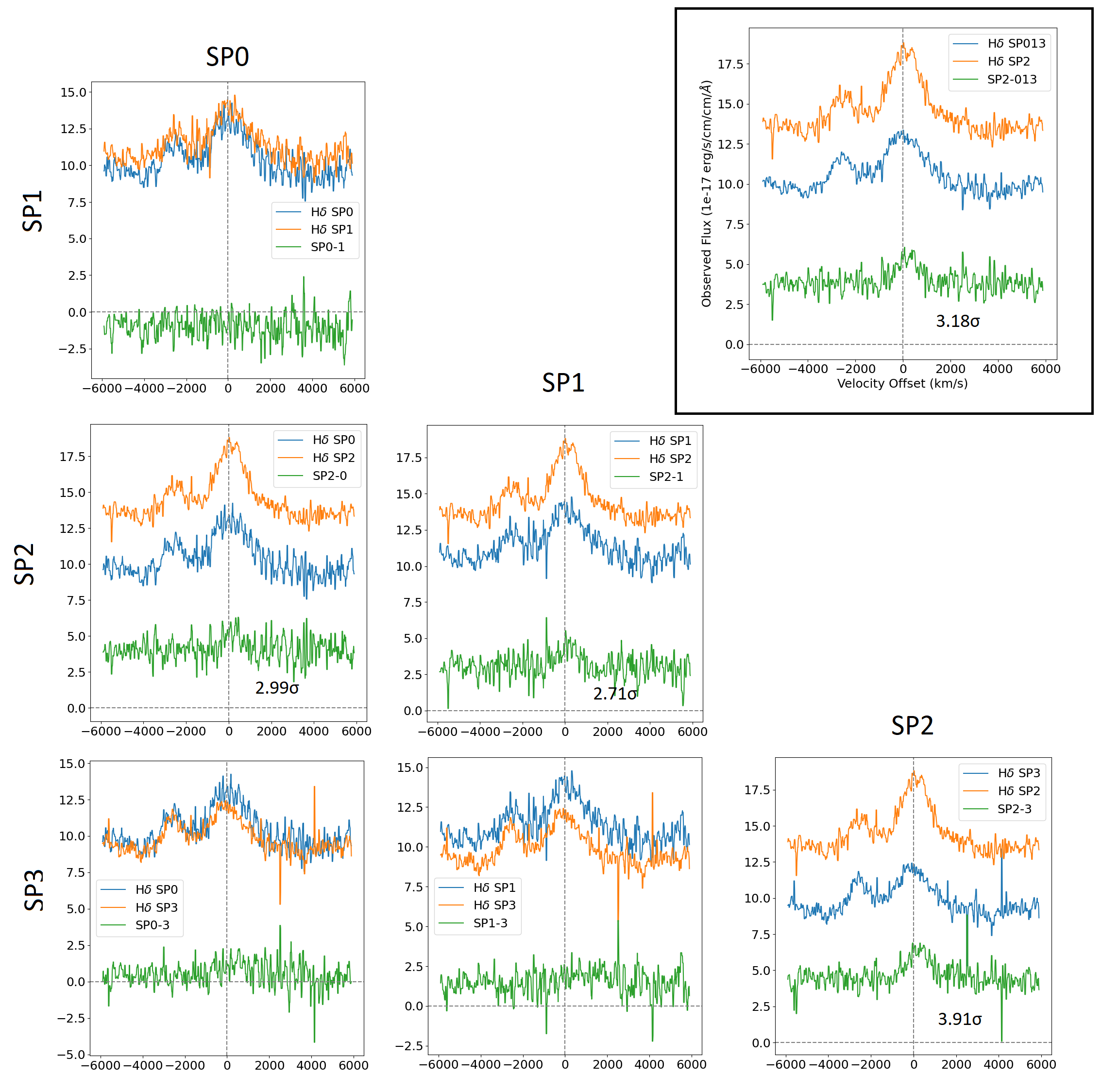}
    \caption{Similar plots to Figure~\ref{fig:excess}, showing all of the lines with enhancements or red-excess. }
\end{figure*}

\begin{figure*}\ContinuedFloat
    \includegraphics[width=0.95\textwidth]{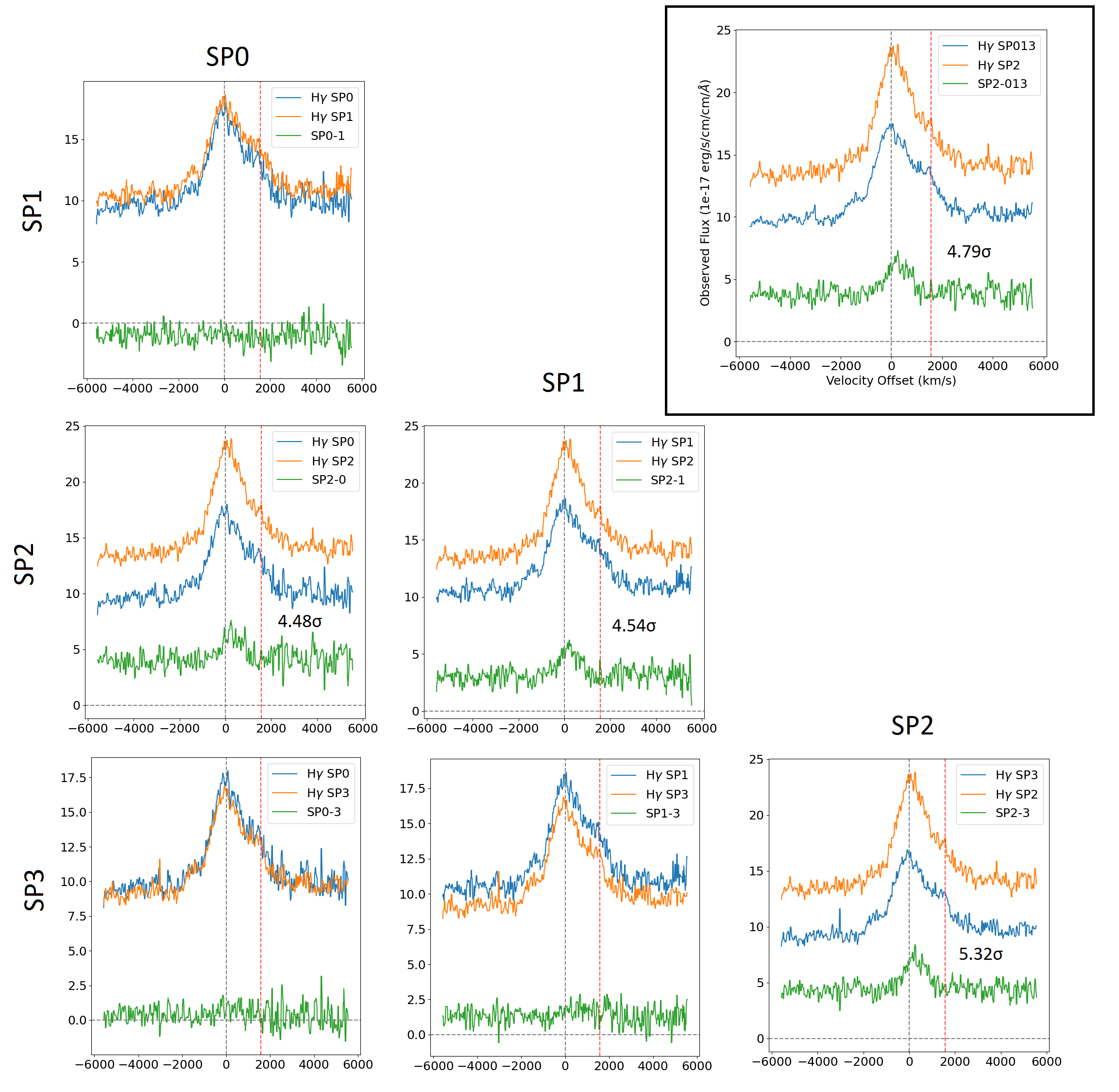}
    \caption{Continued. For the grid plots for H$\gamma$, the additional vertical red dashed line indicates the wavelength position of [\ion{O}{iii}]$\lambda$4364, and they appear to be properly cancelled in the subtracted line for all.}
\end{figure*}

\begin{figure*}\ContinuedFloat
    \includegraphics[width=0.95\textwidth]{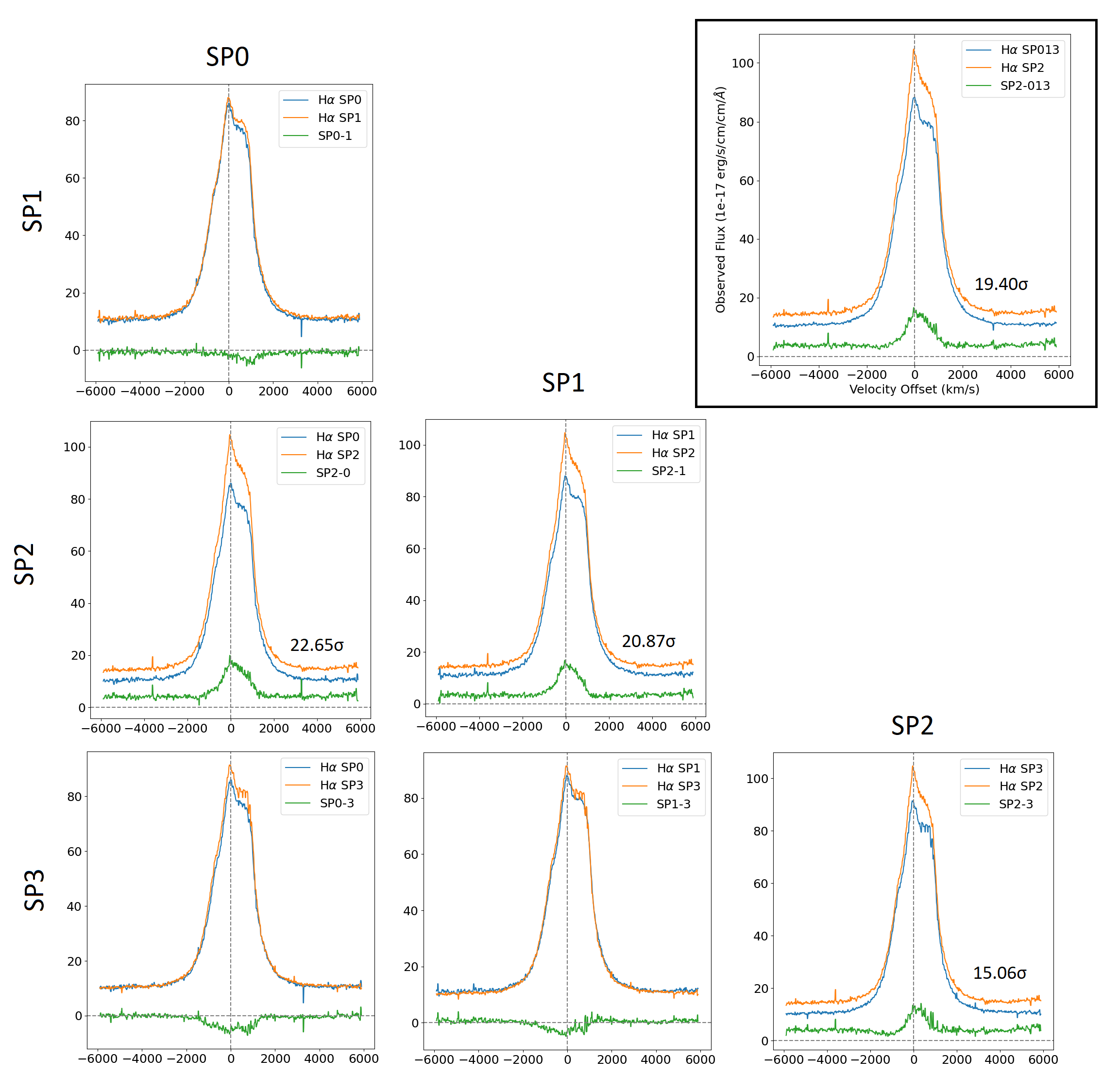}
    \caption{Continued.}
\end{figure*}

\begin{figure*}\ContinuedFloat
    \includegraphics[width=0.95\textwidth]{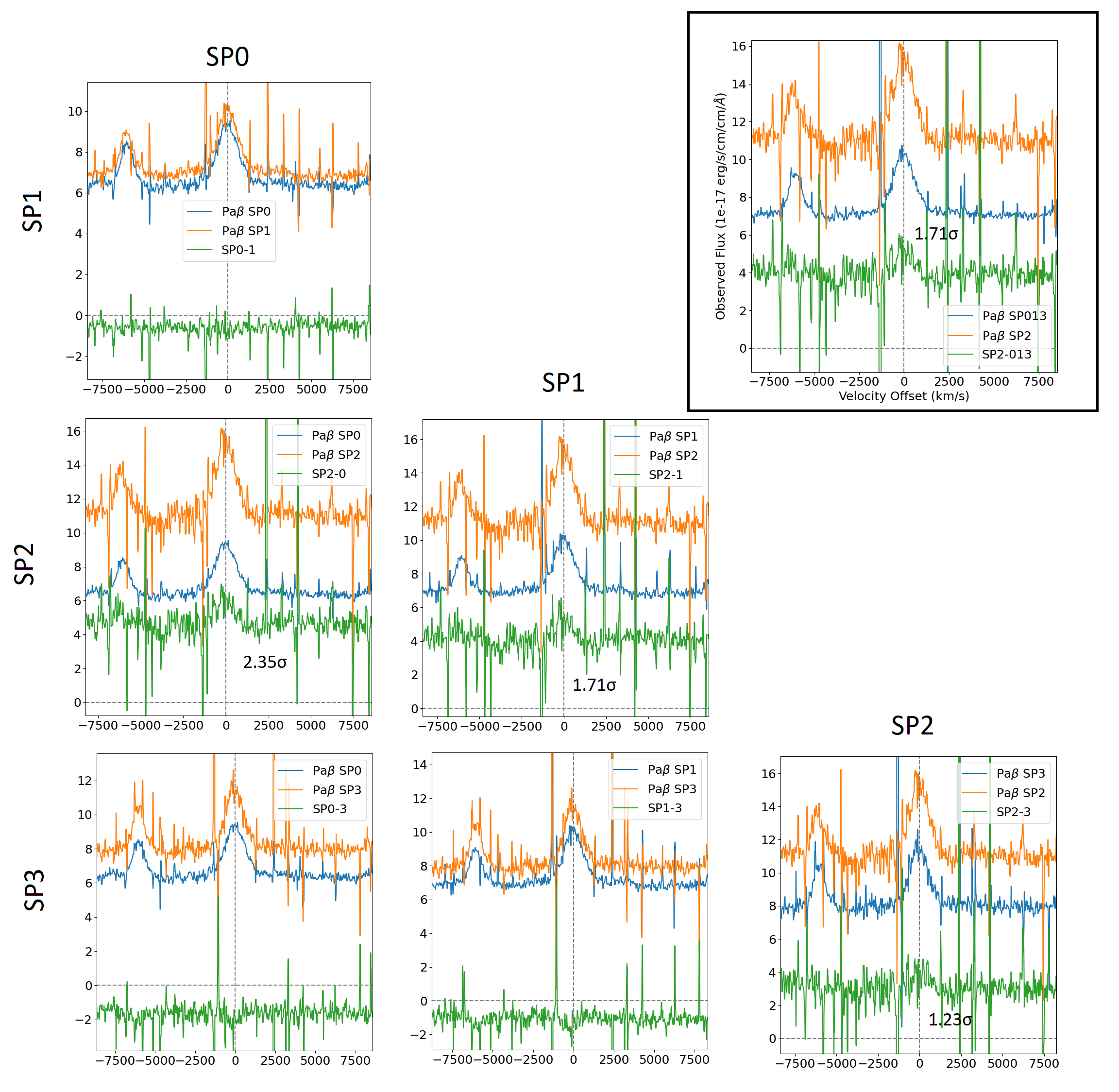}
    \caption{Continued. Statistical significance measured using RMS value taken from SP2 featureless continuum between 13000-13500\AA.}
\end{figure*}

\begin{figure*}\ContinuedFloat
    \includegraphics[width=0.95\textwidth]{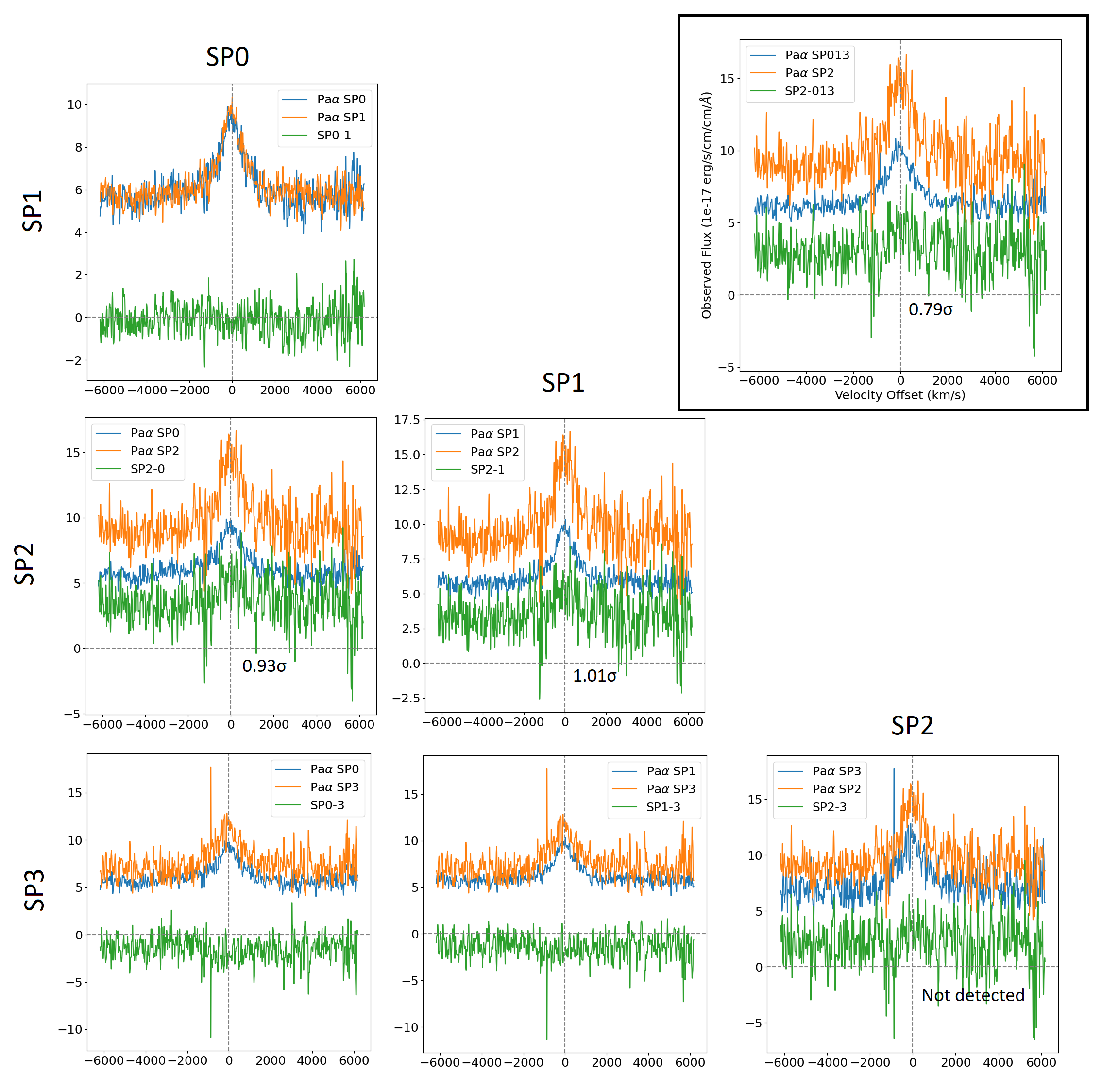}
    \caption{Continued. Statistical significance measured using RMS value taken from SP2 featureless continuum between 18000-18500\AA. This is different to Pa$\beta$ as the quality from X-Shooter degrades towards the red end of the IR arm.}
\end{figure*}

\begin{figure*}\ContinuedFloat
    \includegraphics[width=0.95\textwidth]{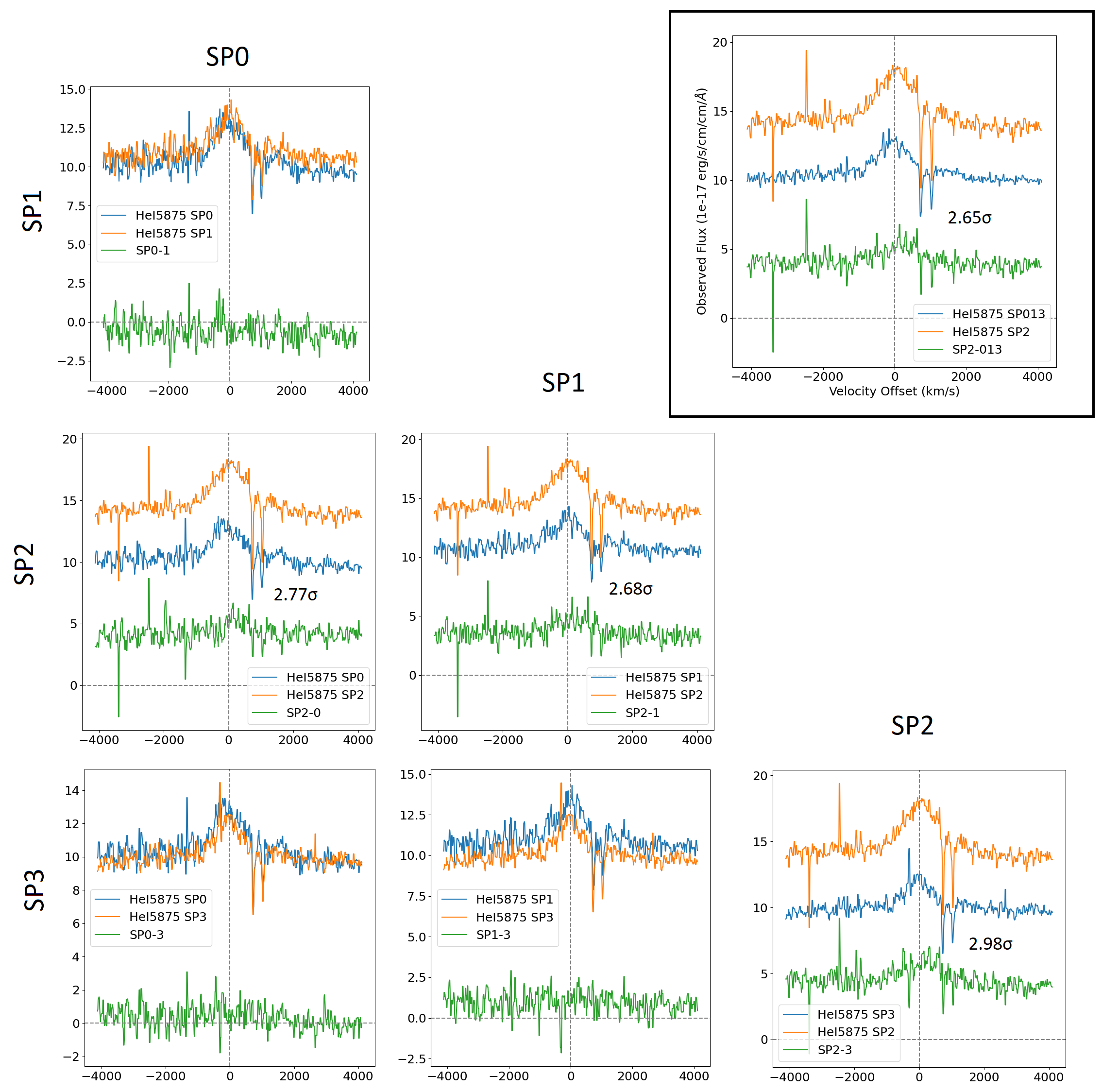}
    \caption{Continued.}
\end{figure*}

\begin{figure*}\ContinuedFloat
    \includegraphics[width=0.95\textwidth]{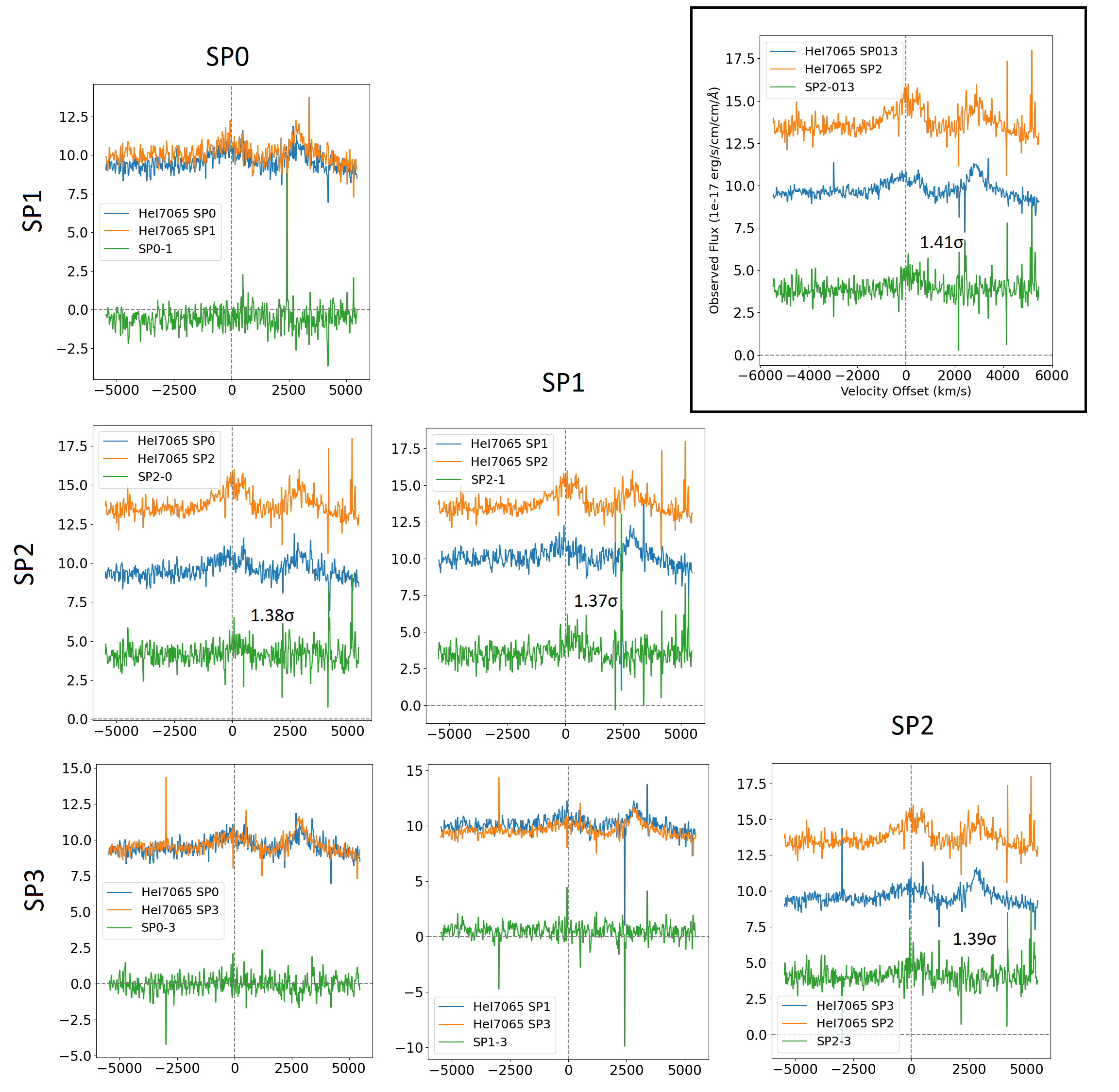}
    \caption{Continued.}
\end{figure*}

\begin{figure*}\ContinuedFloat
    \includegraphics[width=0.95\textwidth]{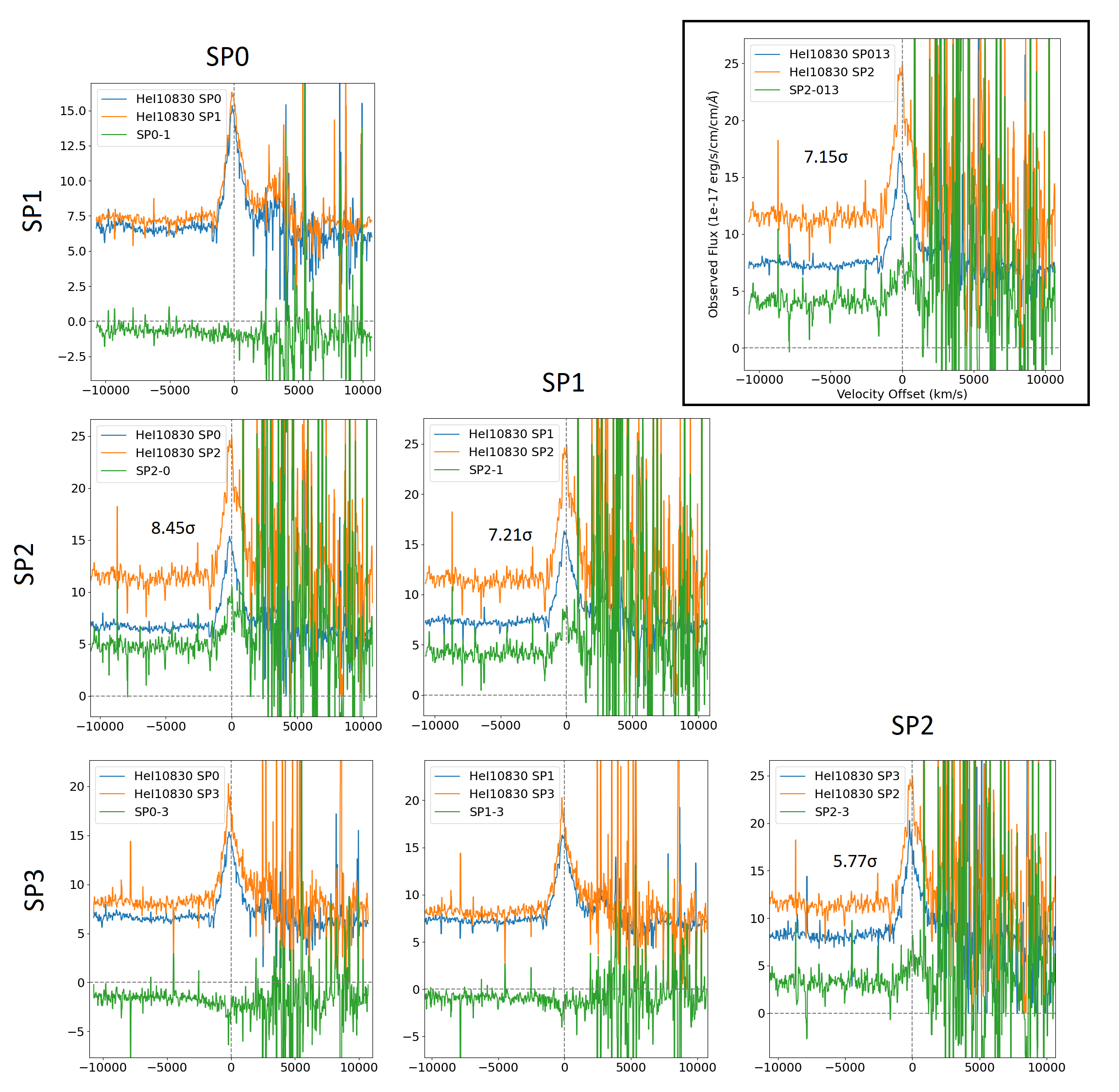}
    \caption{Continued. Statistical significance measured using RMS value taken from SP2 featureless continuum between 13000-13500\AA.}
    \label{appdx:excess}
\end{figure*}

\end{appendix}

\end{document}